\documentclass[aps, twocolumn, pra, superscriptaddress, floatfix]{revtex4-2}
\usepackage{graphicx}
\usepackage{amsmath}
\usepackage{hyperref}
\usepackage[utf8]{inputenc}

\usepackage{color}

\begin{document}

\title{Number-Resolved Detection of Dark Ions in Coulomb Crystals}
\date{\today}

\author{Fabian Schmid}
\email{fabian.schmid@mpq.mpg.de}
\affiliation{Max-Planck-Institut für Quantenoptik, 85748 Garching, Germany}
\author{Johannes Weitenberg}
\affiliation{Max-Planck-Institut für Quantenoptik, 85748 Garching, Germany}
\author{Jorge Moreno}
\affiliation{Max-Planck-Institut für Quantenoptik, 85748 Garching, Germany}
\author{Theodor W. Hänsch}
\affiliation{Max-Planck-Institut für Quantenoptik, 85748 Garching, Germany}
\affiliation{Fakultät für Physik, Ludwig-Maximilians-Universität München, 80799 München, Germany}
\author{Thomas Udem}
\affiliation{Max-Planck-Institut für Quantenoptik, 85748 Garching, Germany}
\affiliation{Fakultät für Physik, Ludwig-Maximilians-Universität München, 80799 München, Germany}
\author{Akira Ozawa}
\email{akira.ozawa@mpq.mpg.de}
\affiliation{Max-Planck-Institut für Quantenoptik, 85748 Garching, Germany}

\begin{abstract}
While it is straightforward to count laser-cooled trapped ions by fluorescence imaging, detecting the number of dark ions embedded and sympathetically cooled in a mixed ion crystal is more challenging. We demonstrate a method to track the number of dark ions in real time with single-particle sensitivity. This is achieved by observing discrete steps in the amount of fluorescence emitted from the coolant ions while exciting secular motional resonances of dark ions. By counting the number of fluorescence steps, we can identify the number of dark ions without calibration and without relying on any physical model of the motional excitation. We demonstrate the scheme by detecting H$_2^+$ and H$_3^+$ ions embedded in a Be$^+$ ion Coulomb crystal in a linear radio frequency trap. Our method allows observing the generation and destruction of individual ions simultaneously for different types of ions. Besides high-resolution spectroscopy of dark ions, another application is the detection of chemical reactions in real time with single-particle sensitivity. This is demonstrated in this work.
\end{abstract}

\maketitle

Trapped and laser-cooled ions have been used to investigate fundamental light-atom interactions~\cite{itano2015, eichmann1993, stalgies1996, eschner2001, leibfried2003}, for observation of ion-neutral chemical reactions~\cite{drewsen2003, roth2006a, willitsch2008, yang2018, tomza2019}, as well as for optical frequency standards~\cite{diddams2001, huntemann2016} and quantum computing~\cite{blatt2008, debnath2016, pogorelov2021}.

Sympathetic cooling is a way for extending this technology to atomic or molecular ions that do not possess suitable transitions for laser cooling. In this scheme, the ions are trapped together with another ion species that can be laser-cooled. Due to the mutual Coulomb interaction, the ions rapidly thermalize, indirectly cooling all species. At sufficiently low temperatures, the ions form regular Coulomb crystals in the trap~\cite{hornekaer2001}. The technique has found application in precision spectroscopy of HD$^+$ and H$_2^+$ molecular ions~\cite{blythe2005, alighanbari2020, patra2020, schmidt2020b, kortunov2021}, optical atomic clocks based on quantum logic spectroscopy of Al$^+$~\cite{schmidt2005, brewer2019}, spectroscopy of highly-charged ions~\cite{kozlov2018, micke2020}, and the study of chemical reactions with molecular ions~\cite{baba2002a, roth2006a, willitsch2008a, hall2012, okada2013}.

While the number of laser-cooled ions can be easily measured by fluorescence imaging, identification and counting of the non-fluorescing dark ions is more difficult. One method is to eject the ions from the trap and to accelerate them onto a detector in an electric field. The different ion species can be distinguished by their arrival times~\cite{schneider2014, meyer2015, rosch2016, schmid2017}. While this method allows a quantitative measurement of the number of ions of each species, it has the disadvantage that it is destructive and a new ion crystal has to be prepared after each measurement.

In linear Paul traps, lighter ions are more tightly confined than heavier ions. Lighter sympathetically cooled ions therefore form a dark region in the center of the fluorescence image of an ion crystal consisting of a heavier coolant species. The number of dark ions can then be obtained by comparing experimental images with simulated ones~\cite{blythe2005, germann2014, okada2015a}. This method is non-destructive, and the ion images can be acquired quickly and post-processed later. However, different dark ion species cannot be distinguished.

Instead, secular excitation has been used for non-destructive detection of trapped ions. In this method the secular motion of the ions, i.e.\ the harmonic motion in the time averaged trap potential, is excited resonantly by applying an additional oscillating electric field. This transfers energy into the motion of the surrounding coolant ions and thereby increases their temperature. Due to the temperature dependence of the Doppler broadening this leads to a change in the amount of fluorescence that can be observed from the coolant ions. The secular motion of the ions in three-dimensional Coulomb crystals has rich dynamics that can complicate the analysis of the secular excitation spectra. For example, the frequencies of the secular resonances are influenced by space charge effects and the mechanical coupling between the ions~\cite{roth2007, zhang2007, ann2019}. The energy transfer to the coolant ions is expected to increase with an increasing number of dark ions. Therefore, the fluorescence change induced by motional excitation serves as a measure of the number of dark ions. However, the relationship between the fluorescence change and the number of excited dark ions is in general non-linear and is influenced by various experimental parameters, such as the strength of the motional excitation and the geometry of the mixed ion crystal. Therefore, evaluating the number of dark ions quantitatively is challenging and often requires intricate modeling and calibration of the signal using molecular dynamics simulations~\cite{baba2002a, Biesheuvel2016}. This problem has been limiting the usage of the secular excitation method for highly precise spectroscopy so far.

In this work, we show that by properly choosing experimental parameters, discrete steps in the secular excitation signal can be observed that are identified with individual dark ions leaving the trap or being generated within the trap. Hence such a signal is auto-calibrating and counting the number of ions gives the ultimate accuracy. The signal does not have to be calibrated using a physical model of the secular excitation. Spurious signals at other frequencies that may arise from motional coupling have no influence on the counting process and can be safely ignored.

We experimentally demonstrate this method by resonantly exciting the radial motion of H$_2^+$ and H$_3^+$ ions embedded in a laser-cooled Be$^+$ ion crystal. We observe concomitant changes in the amount of fluorescence from the Be$^+$ ions when the number of trapped H$_2^+$ or H$_3^+$ ions changes due to chemical reactions with neutral rest gas molecules.

Spectroscopy of dark ions requires a scheme for detecting that the target transition is being excited. This can for example be achieved by monitoring that new ion species are created by state-dependent photoionization~\cite{herrmann2009a} or resonance-enhanced multiphoton dissociation~\cite{Biesheuvel2016, alighanbari2020, patra2020, schmidt2020b}. The reliable detection of single dark ions demonstrates that this detection scheme can be single-event sensitive and that the spectroscopy will be ultimately limited only by quantum projection noise. Nonlinearities in the signal intensity may introduce a systematic frequency shift, especially when the spectrum consists of multiple overlapping lines~\cite{Biesheuvel2016}. Accurate counting of the dark ions gives rise to a spectroscopy signal with negligible nonlinearity.

Chemical reactions at ultracold temperatures can be investigated precisely for a small number of atoms or ions after careful quantum-state preparation~\cite{ratschbacher2012}. Our detection scheme can be employed to efficiently capture such events with single-particle sensitivity.

\begin{figure}[htb]
    \centering
    \includegraphics[scale=.3]{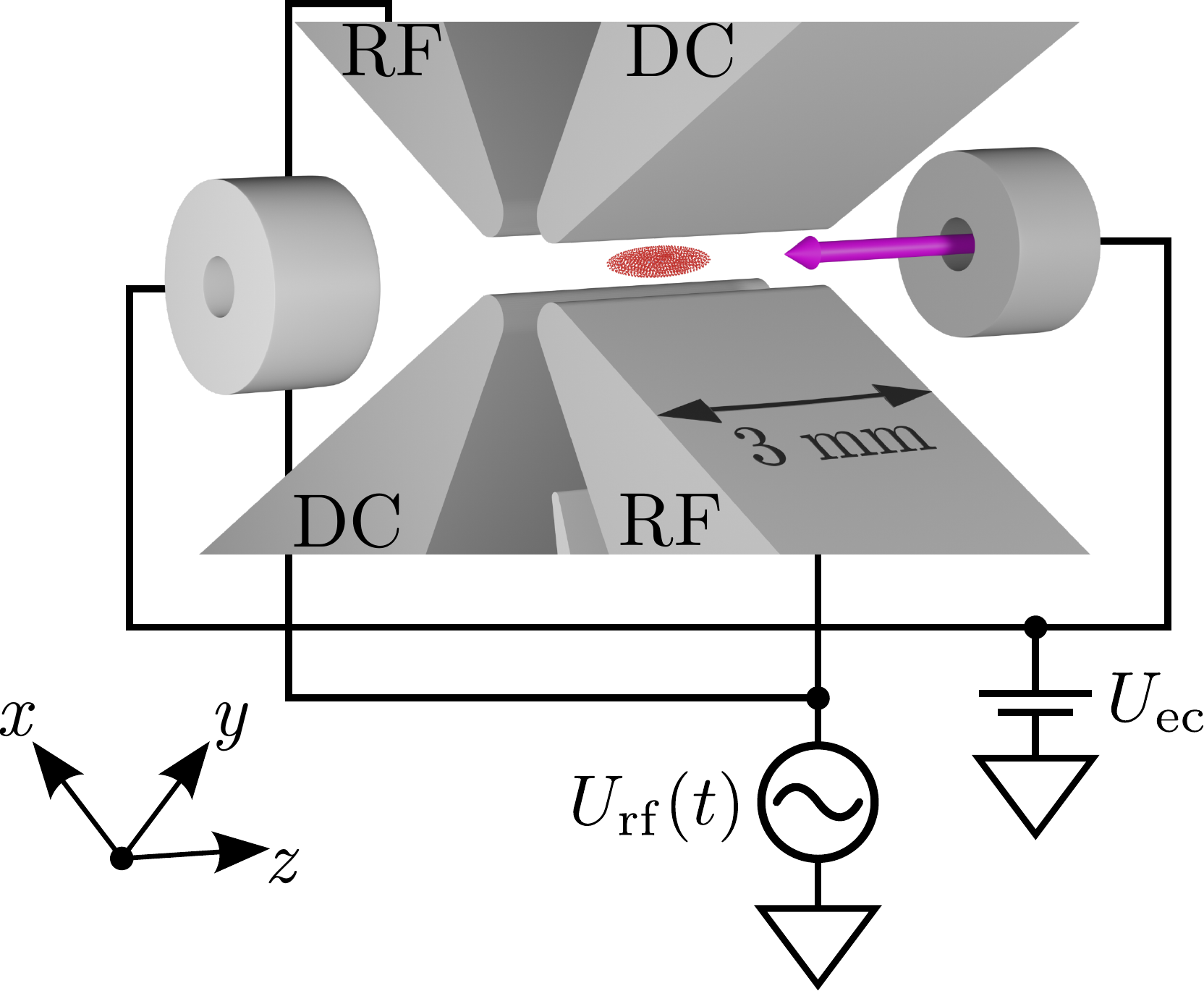}

    \vspace{.2cm}

    \includegraphics[scale=.3]{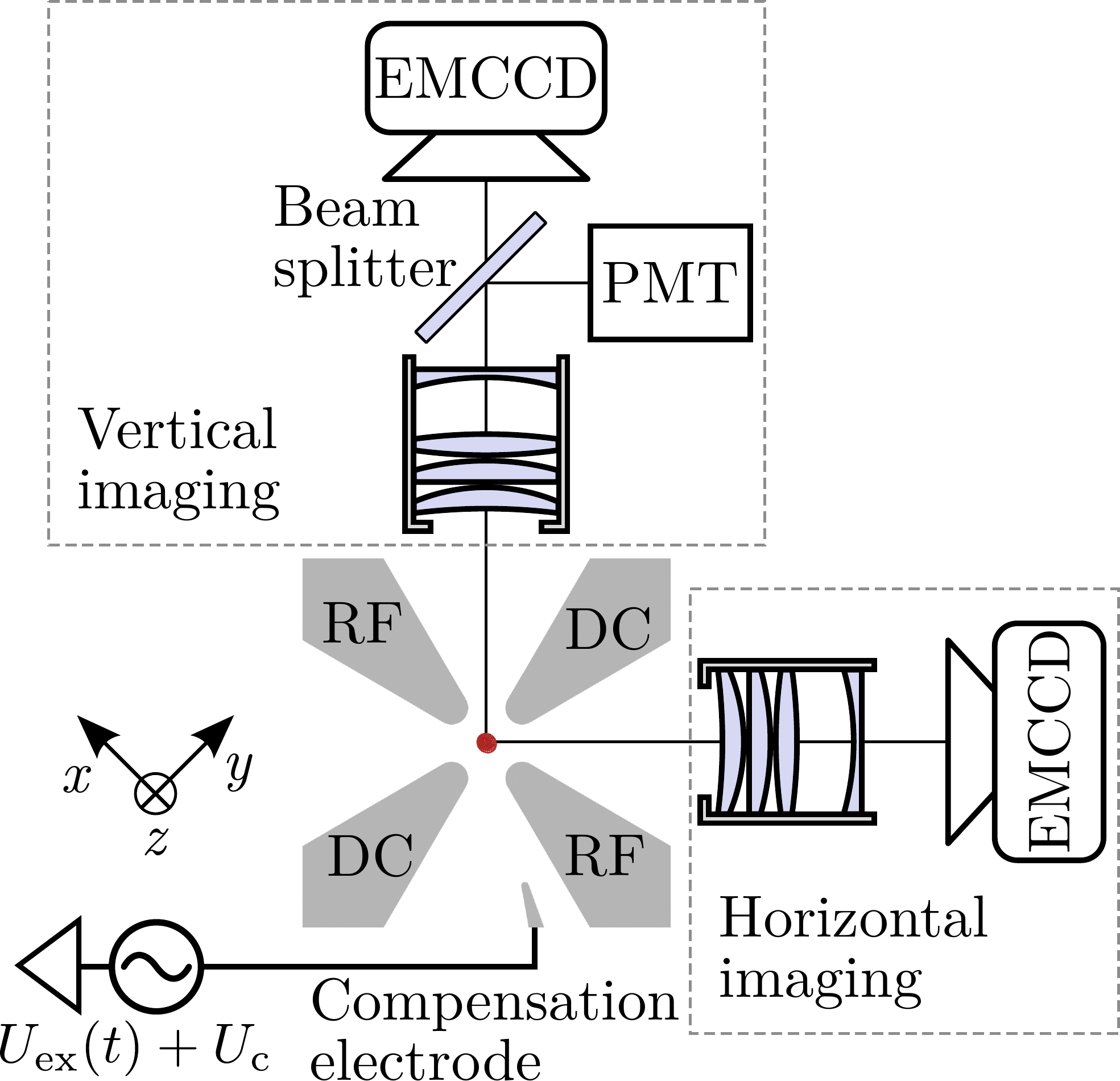}
    \caption{Geometry of the ion trap setup. Radial (top) and axial (bottom) view. The Coulomb crystal is not drawn to scale. Radial confinement is generated by the RF voltage $U_\mathrm{rf} (t)$ that is applied to one diagonal pair of blade electrodes. A static voltage $U_\mathrm{ec}$ is applied to the endcap electrodes for axial confinement. The cooling beam (purple) is directed along the trap axis ($z$). The fluorescence from the laser-cooled Be$^+$ ions is imaged onto two EMCCD cameras. The static compensation voltage $U_\mathrm{c}$ and voltages applied to the two dc blade electrodes are used to compensate stray fields in the trap caused by patch potentials and electrode misalignment. A sinusoidal voltage $U_\mathrm{ex} (t)$ is added to the compensation electrode in order to excite the secular motion of the trapped ions.}%
    \label{fig:experimental_setup}
\end{figure}

We use a linear Paul trap to confine the ions. As shown in Fig.~\ref{fig:experimental_setup}, it consists of four blade electrodes that are spaced 0.45~mm from the trap axis and have an axial length of 3.00~mm. Axial confinement is provided by two endcap electrodes that are located 3.50~mm from the trap center.

The trap is driven with a radio frequency (RF) of 66.05~MHz with an amplitude of around 120~V, and a static voltage of 400~V is applied to the endcaps. For Be$^+$ this results in a radial secular frequency of around 1.6~MHz, corresponding to a Mathieu stability parameter $q \approx 0.07$~\cite{leibfried2003}, and an axial secular frequency of 645~kHz. The single-particle radial secular frequency scales proportional to the ion's charge-to-mass ratio. The theoretical values are 4.8~MHz and 7.2~MHz for H$_3^+$ and H$_2^+$, respectively.

The Be$^+$ ions are laser-cooled by driving the 2s~$^2$S$_{1/2}$ ($F$=2) $\rightarrow$ 2p~$^2$P$_{3/2}$ ($F$=3) cycling transition with circularly polarized 313~nm light that is generated by sum-frequency generation of two continuous wave fiber lasers at 1051~nm and 1550~nm and subsequent frequency-doubling~\cite{wilson2011}. The cooling transition has a natural linewidth of $\Gamma = 2\pi \times 18$~MHz~\cite{fuhr2010}, and the cooling beam contains two frequency components that are red-detuned from the transition by 130~MHz and 460~MHz. We found that adding the far-detuned component makes the system more robust against losing the trapped ions when strongly driving secular excitations. Both frequency components can be switched on or off separately and have similar intensities of around $I_{\mathrm{sat}}$, where $I_{\mathrm{sat}} = 765~\textup{W}/\textup{m}^2$ is the saturation intensity of the cooling transition~\cite{fuhr2010}. The cooling beam is aligned parallel to the trap axis and propagates through holes in the end-cap electrodes. A weak magnetic field is applied parallel to the beam in order to define the quantization axis. We use an electro-optic modulator to create 1.25~GHz sidebands on the cooling laser in order to re-pump the ions out of the 2s~$^2$S$_{1/2}$ ($F$=1) dark state.

The trap is housed in an ultra-high vacuum chamber. The background pressure measured with a cold-cathode ion gauge is around 4~$\times$ 10$^{-11}$ mbar.

We load Be$^+$ ions into the trap with the help of a beam of beryllium atoms from an oven. The beam is sent through the trapping region where it is overlapped with a 235~nm laser beam that resonantly ionizes the atoms~\cite{lo2013}. During Be$^+$ loading, initially only the far-detuned frequency component is applied to efficiently cool the hot ions. Once fluorescence from trapped ions is detected, the near-detuned frequency component is turned on. This strongly increases the fluorescence signal, such that the ion crystal shape can be monitored. The ionization laser is turned off when the desired ion crystal size is reached.

We then turn on a home-built electron gun for 1~s with a current of about 95~µA. The electrons ionize some of the hydrogen molecules from the residual gas in our vacuum chamber, and the resulting molecular ions become embedded in the ion crystal. 

As shown in Fig.~\ref{fig:experimental_setup}, we use two imaging systems to observe the fluorescence from the laser-cooled ions from the horizontal and vertical direction perpendicular to the trap axis. In each system, the light is imaged onto an electron-multiplying CCD (EMCCD) camera using a home-built objective with magnification M $\approx 9$ and a numerical aperture of 0.23. The vertical imaging system additionally contains a beam splitter that sends roughly half of the light onto a photomultiplier tube (PMT).

\begin{figure}[htb]
    \centering
    \includegraphics[width=\linewidth]{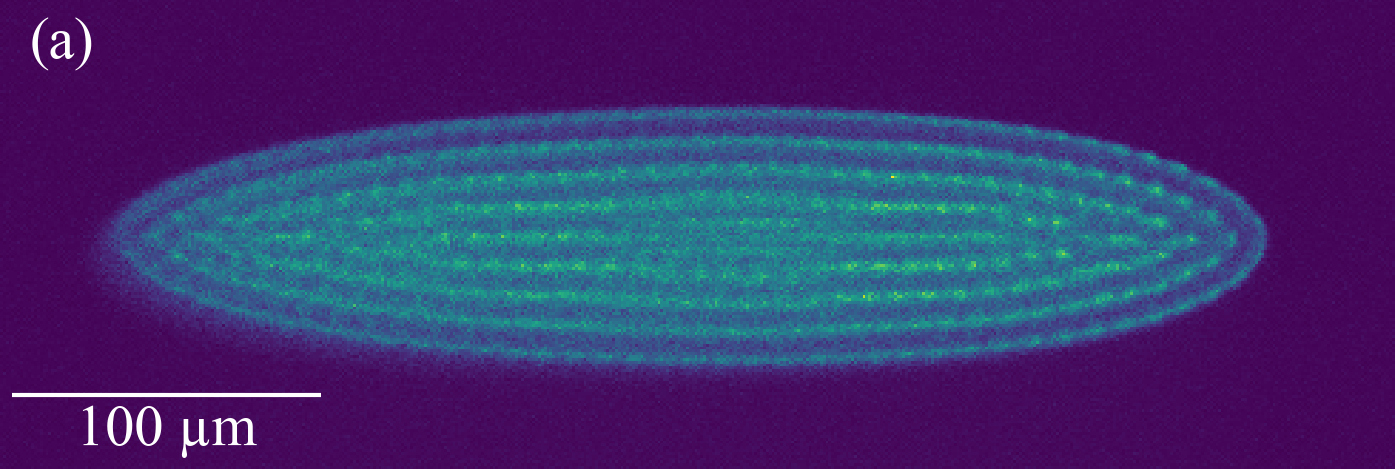}
    \includegraphics[width=\linewidth]{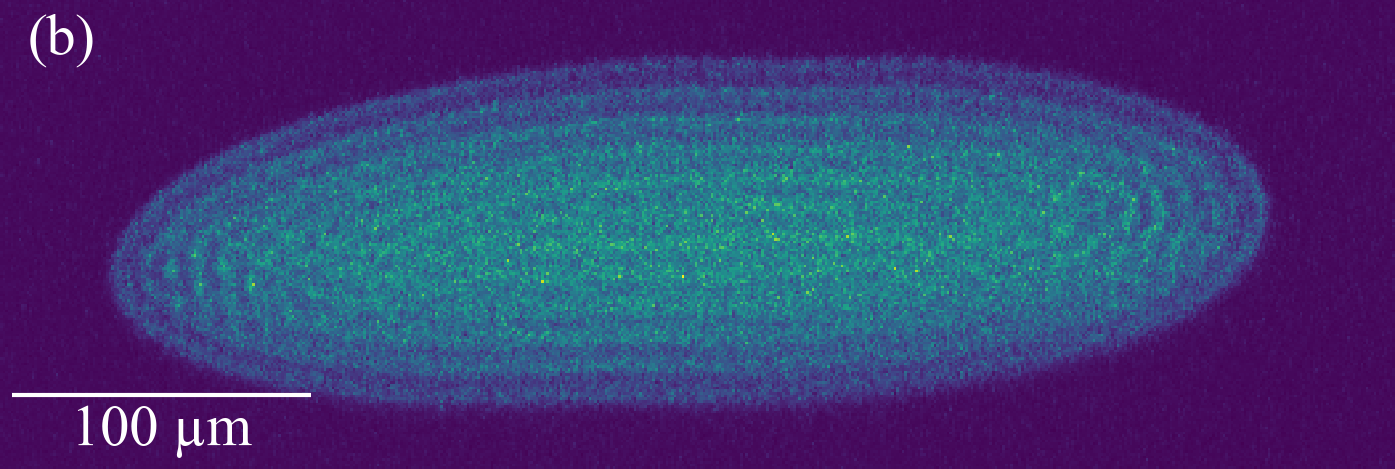}
    \includegraphics[width=\linewidth]{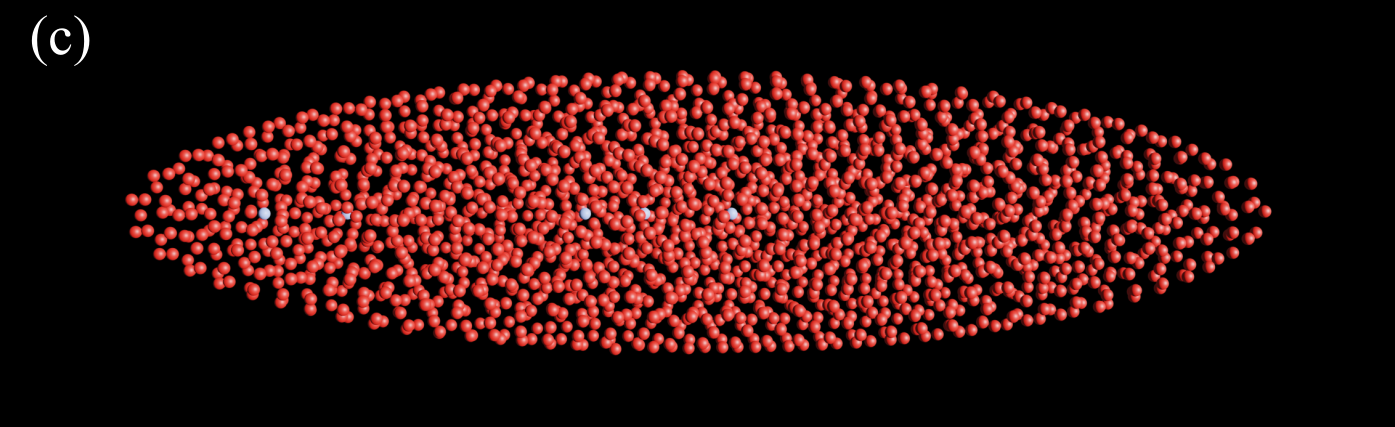}
    \includegraphics[width=\linewidth]{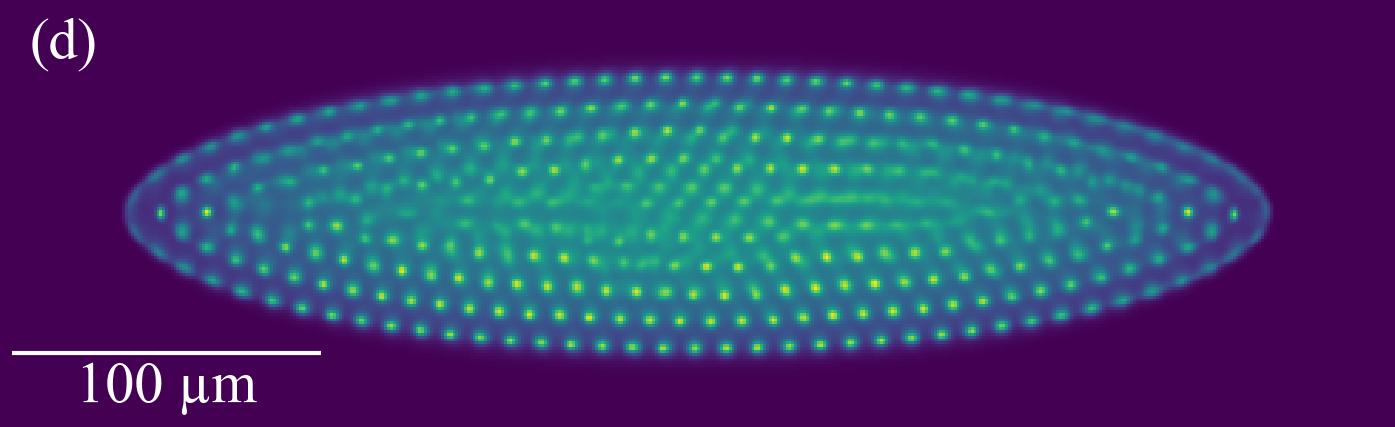}
    \caption{Fluorescence images of an ion crystal consisting of around 2000 Be$^+$ ions with a few embedded H$_2^+$ and H$_3^+$ ions, horizontal (a) and vertical (b) view. (c) Coulomb crystal containing 2000 Be$^+$ ions (red spheres) and 5 H$_2^+$ ions (light blue spheres) obtained by the molecular dynamics simulation. (d) Simulated ion image for the crystal shown in (c) at 10~mK.}%
    \label{fig:ion_crystal_image}
\end{figure}

Fig.~\ref{fig:ion_crystal_image}(a) and (b) show typical fluorescence images of an ion crystal in our setup. We believe that the asymmetric shape visible in Fig.~\ref{fig:ion_crystal_image}(b) is due to imperfect alignment of the trap electrodes.

\begin{figure}[htb]
    \centering
    \includegraphics[width=\linewidth]{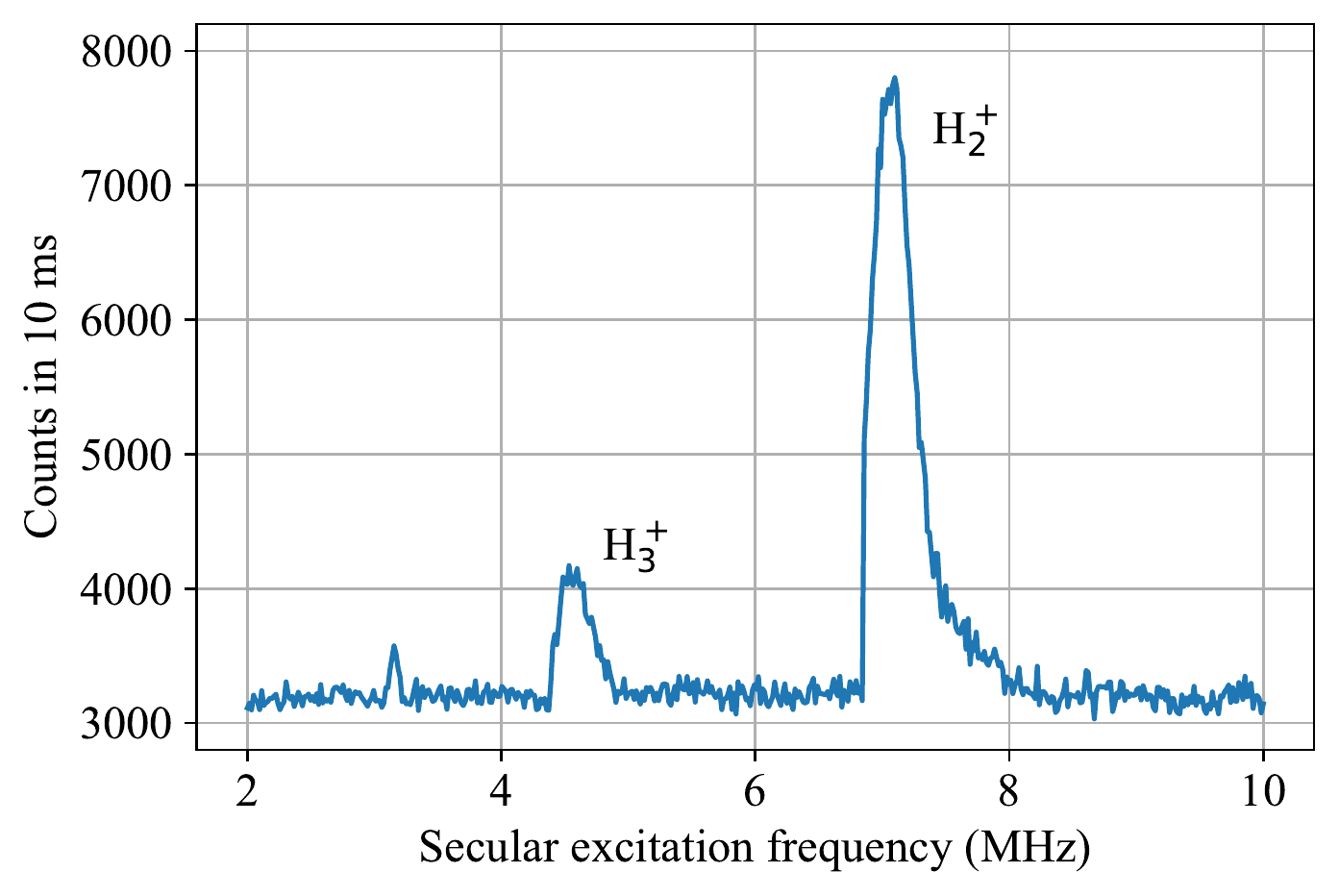}
    \caption{Secular spectrum showing the presence of H$_2^+$ and H$_3^+$ ions embedded in the Be$^+$ coolant ions.}%
    \label{fig:secular_spectrum}
\end{figure}

In order to measure the radial secular frequencies of the trapped ions, we excite their motion by applying a sinusoidal voltage to the compensation electrode that is located close to the trap electrodes (see Fig.~\ref{fig:experimental_setup}). We sweep the frequency range between 2~MHz and 10~MHz in 500 steps and for each frequency collect photon counts with the PMT for 10~ms. This is slow enough that the ion motion reaches steady state for each frequency point. Fig.~\ref{fig:secular_spectrum} shows a typical resulting secular spectrum. Two peaks at 4.6~MHz and 7.1~MHz can be observed which we attribute to H$_3^+$ and H$_2^+$ ions, respectively. The electrode that we use for the secular excitation is low-pass filtered with a 1~nF capacitor and is not impedance matched to the 50~$\Omega$ output impedance of the function generator that produces the excitation signal. This makes the electric field amplitude at the position of the ions frequency dependent. In addition, the efficiency of the motional excitation might be different for H$_3^+$ and H$_2^+$ ions. We therefore use two different excitation amplitudes for the frequency range from 2~MHz to 5.5~MHz and from 5.5~MHz to 10~MHz. The amplitudes are chosen to give roughly the same signal strength per ion for the H$_2^+$ and H$_3^+$ resonances.

\begin{figure}[htb]
    \centering
    \includegraphics[width=\linewidth]{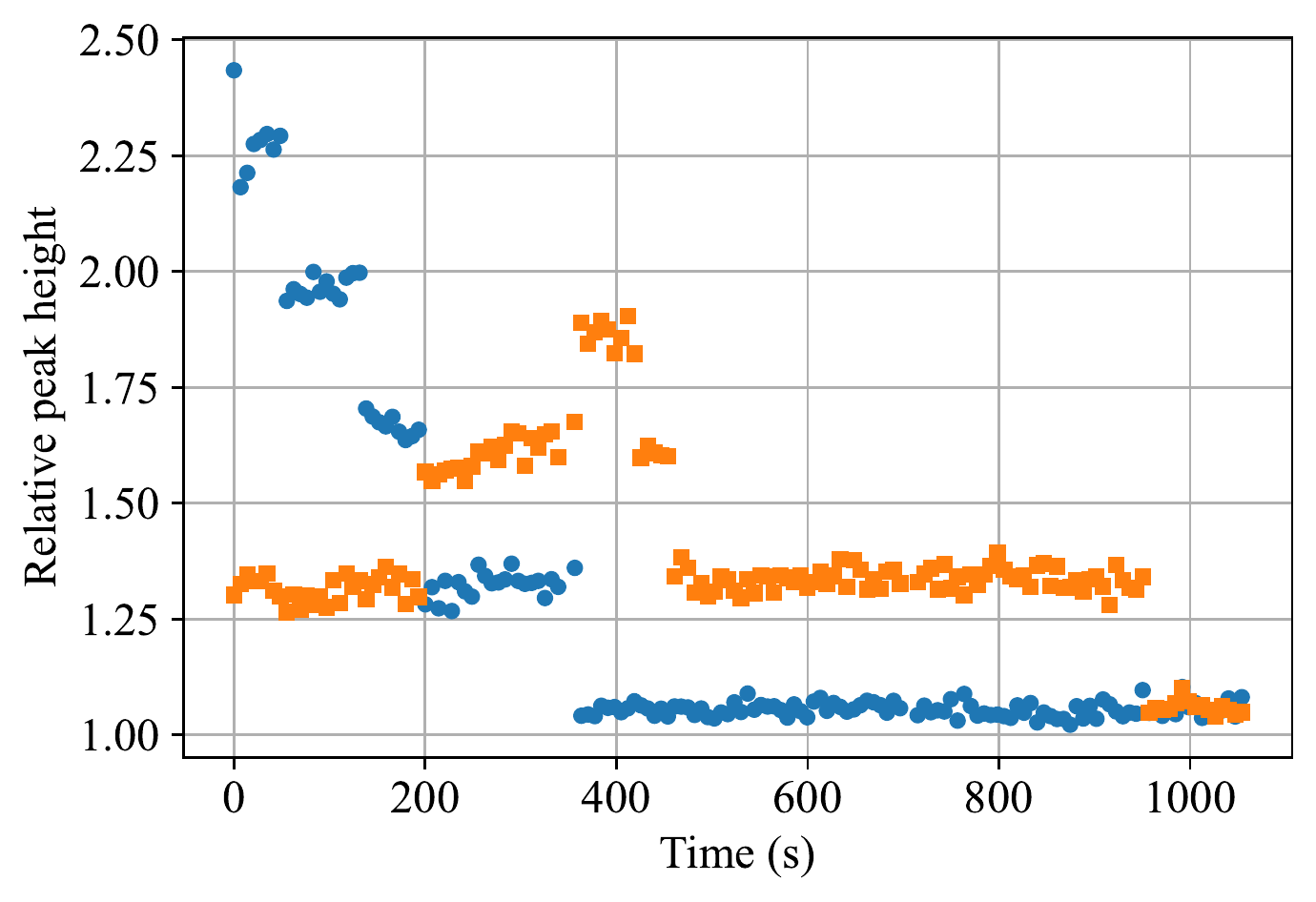}
    \caption{Change of the fluorescence peak heights at the H$_2^+$ (blue circles) and H$_3^+$ (orange squares) resonances over time. For each measurement, the peak heights are normalized to the off-resonant scattering rate. The stepwise changes of the peak heights are caused by individual ions undergoing chemical reactions with residual gas molecules. Within each step, the peak heights scatter by much less than the step-to-step difference. Each secular scan therefore allows a reliable determination of the ion numbers.}%
    \label{fig:time_evolution}
\end{figure}

We then repeatedly perform secular scans over a period of a few minutes. Fig.~\ref{fig:time_evolution} shows how the height of the secular resonances changes over time in a typical experiment. We normalize the peak heights of each scan to the off-resonant scattering rate. This compensates for small drifts in the cooling laser power and Be$^+$ ion loss during the measurement. The height of the H$_2^+$ signal drops in steps until after around 360~s no peak is visible anymore. The height of the H$_3^+$ signal increases in steps that coincide with drops of the H$_2^+$ signal. We attribute this to individual H$_2^+$ ions reacting with molecular hydrogen from the residual gas to form neutral hydrogen and H$_3^+$ according to the exothermic chemical reaction H$_2^+$ + H$_2 \rightarrow $ H$_3^+$ + H~\cite{bowers1969}. The clearly visible steps in the signals show that we are observing individual ions being destroyed or created within the Coulomb crystal. Sometimes the H$_2^+$ signal drops without a corresponding increase in H$_3^+$ signal. One possible explanation is that the H$_2^+$ ion reacted with a different residual gas molecule such as N$_2$, O$_2$, or H$_2$O~\cite{kim1975a}. Another possibility is that the kinetic energy transferred to the H$_3^+$ molecule in the chemical reaction is large enough to sometimes allow the product to escape from the trap. Rate constants for the chemical reactions between H$_2^+$ and typical residual gas molecules are in the range of 2--7$\times 10^{-9}$~cm$^3$ s$^{-1}$~\cite{bowers1969, kim1975a}. An exponential fit to the H$_2^+$ ion number yields a total decay rate of 0.006~s$^{-1}$ which corresponds to a residual gas pressure in the range of $4 \times 10^{-11}$ mbar to $1 \times 10^{-10}$ mbar, in rough agreement with the pressure indicated by the vacuum gauge. The number of H$_3^+$ ions subsequently drops in similar steps over the timescale of a few minutes. We attribute this to proton transfer reactions of the type H$_3^+$ + X $\rightarrow$ HX$^+$ + H$_2$ which are possible with a number of residual gas molecules such as N$_2$, CO, CO$_2$, and H$_2$O~\cite{burt1970}.

\begin{figure}[htb]
    \centering
    \includegraphics[width=\linewidth]{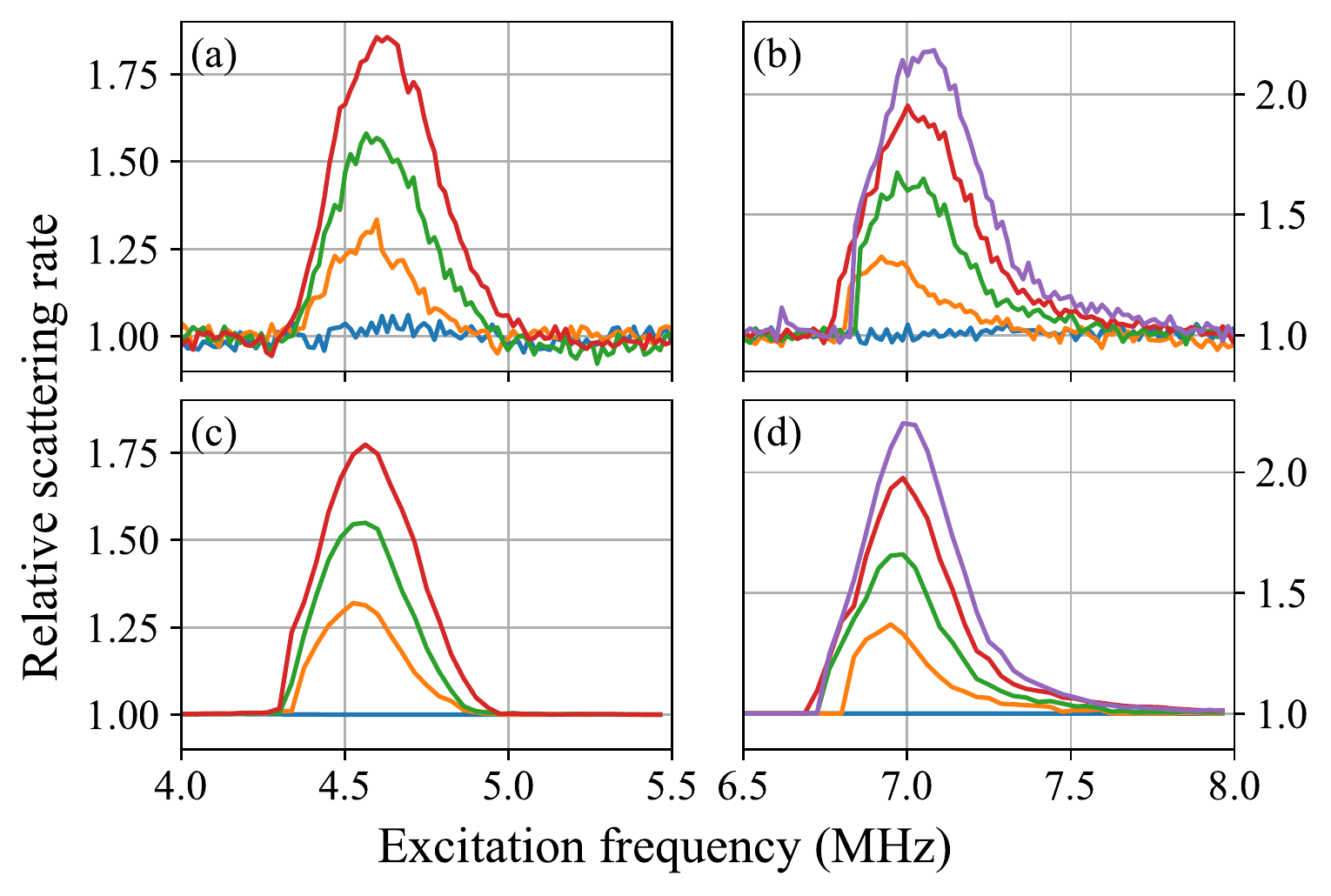}
    \caption{Experimental secular excitation spectra for (a) 0--3 H$_3^+$ ions and (b) 0--4 H$_2^+$ ions. (c) and (d) are spectra obtained from molecular dynamics simulations with 2000 Be$^+$ ions and the same numbers of H$_2^+$ and H$_3^+$ ions as in the experiment. The simulated excitation amplitudes are 2.9~V/m for the H$_3^+$ resonance and 2.5~V/m for the H$_2^+$ resonance.}%
    \label{fig:secular_spectra_experiment_simulation}
\end{figure}

One striking feature of Fig.~\ref{fig:time_evolution} and Fig.~\ref{fig:secular_spectra_experiment_simulation} is that the fluorescence peak heights are nearly linear in the number of dark ions. We use the LAMMPS molecular dynamics code~\cite{thompson2022, bentine2020} to further investigate this behavior. We first simulate the stationary configuration of the mixed ion crystals without applying the excitation as shown in Fig.~\ref{fig:ion_crystal_image}(c). We generate simulated fluorescence images, taking into account the motion of the ions and the spot size and depth of field of the imaging objective. We then estimate the number of trapped ions in the experiment and their temperature by comparing the simulated images with the experimental ones~\cite{zhang2007, germann2014, okada2015a}. A good agreement is obtained between experimental and simulated images for around 2000 Be$^+$ ions at 10~mK mixed with a few dark ions as shown in Fig.~\ref{fig:ion_crystal_image}(d). In order to simulate the secular excitation spectra, we add an oscillating electric field perpendicular to the trap axis. The amount of fluorescence from the coolant ions is calculated during the motional excitation. The simulation is repeated for different excitation frequencies and numbers of dark ions. Fig.~\ref{fig:secular_spectra_experiment_simulation}(a) and (b) show experimental secular excitation spectra for different numbers of H$_3^+$ and H$_2^+$ ions, and Fig.~\ref{fig:secular_spectra_experiment_simulation}(c) and (d) show the corresponding simulated spectra. The stepwise change of the fluorescence peak height for different numbers of ions is reproduced in the simulations. Both the simulation and the measurement suggest that the change in the amount of fluorescence is roughly proportional to the number of dark ions that are being excited at their secular frequency.

\begin{figure}[htb]
    \centering
    \includegraphics[width=\linewidth]{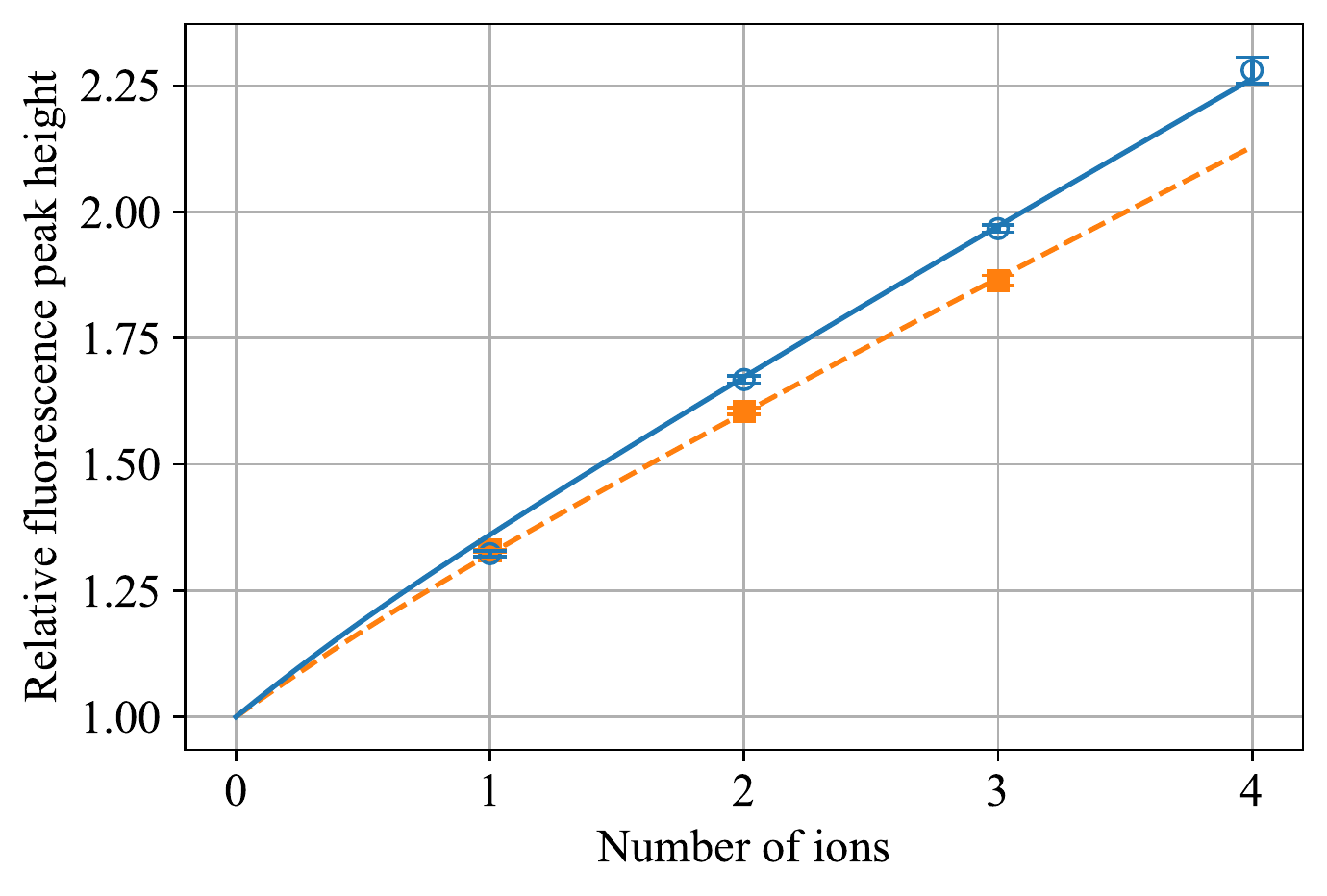}
    \caption{Average peak heights of the H$_2^+$ (blue circles) and H$_3^+$ (orange squares) secular resonances. The error bars show the standard error of the mean. The solid and dashed lines are fits of our model curve $R(T_{\mathrm{s}}) /R(T_0)$ to the H$_2^+$ and H$_3^+$ data, respectively. Separate fits are performed for the H$_2^+$ and H$_3^+$ resonances since the secular excitation strength is different for the two species and two different fit parameters $\alpha_{\mathrm{H}_2^+}$ and $\alpha_{\mathrm{H}_3^+}$ were necessary.}
    \label{fig:peak_heights_theory_fit}
\end{figure}

We now derive a model for the dependence of the peak scattering rate on the number of ions that are being excited. For a cycling cooling transition in an atom at rest the photon scattering rate is given by
\begin{equation}
    w(\Delta)=\frac{s \Gamma/2}{1 + s + (\frac{\Delta}{\Gamma/2})^2},
\end{equation}
where $\Delta$ is the laser detuning and $s = I/I_\mathrm{sat}$ is the saturation parameter. During the secular excitation, the temperature of the coolant ions increases to $T_{\mathrm{s}}$, and the scattering rate per Be$^{+}$ ion is given by~\cite{Biesheuvel2016}
\begin{equation}
    R(T_{\mathrm{s}}) = 
    \int_{-\infty}^{\infty} w(\Delta-kv) p(v,T_{\mathrm{s}}) \mathrm{d}v,
    \label{eqn:scattering_rate}
\end{equation}
where $p(v,T)= \sqrt{m_\mathrm{Be}/\left(2 \pi k_\mathrm{B} T\right)}\exp{\left(-m_\mathrm{Be} v^2/(2 k_\mathrm{B} T)\right)}$ is the Maxwell-Boltzmann distribution. $v$ is the velocity of the ion along the cooling laser beam, $k$ is the cooling laser wave number, and $m_\mathrm{Be}$ is the Be$^{+}$ mass.
We assume that the coolant ion temperature $T_{\mathrm{s}}$ during the secular excitation is determined by the balance between the heating due to the motional excitation and the average laser cooling power. This condition is expressed as 
\begin{equation}
    N_\mathrm{Be} p_\mathrm{c}(T_{\mathrm{s}}) = N_\mathrm{ex} p_\mathrm{h},
    \label{eqn:cooling_balance}
\end{equation}
where $N_\mathrm{Be}$ is the number of Be$^+$ ions and $N_\mathrm{ex}$ is the number of dark ions that are being excited. $p_\mathrm{h}$ is the heating power per excited dark ion.  Note that the average cooling power per coolant ion $p_\mathrm{c}(T_{\mathrm{s}})$ is temperature dependent because of the Doppler broadening of the cooling transition
\begin{equation}
    p_\mathrm{c}(T_{\mathrm{s}})= -\hbar k \int_{-\infty}^{\infty} v w(\Delta-kv) p(v,T_{\mathrm{s}}) \mathrm{d}v.
    \label{eqn:cooling_power}
\end{equation}
We confirmed the validity of our assumption of Eq.~(\ref{eqn:cooling_balance}) using the molecular dynamics simulations (see Supplemental Material~\cite{supplement}). The relative fluorescence peak heights of the secular resonances shown in Fig.~\ref{fig:time_evolution} correspond to $R(T_{\mathrm{s}}) /R(T_0)$ where $T_0$ is the equilibrium temperature of the Be$^{+}$ ions
in the absence of heating due to secular excitation. Under our experimental conditions, $T_0$ is around 10~mK. For simplicity we use $T_0=0$~K which introduces a marginal ($\sim 2\%$) difference in the scattering rate.
It is difficult to precisely estimate the heating rate $p_\mathrm{h}$ from the experimental parameters. We therefore treat $\alpha = p_\mathrm{h}/N_\mathrm{Be}$ as a free parameter in our model. We evaluated $R(T_{\mathrm{s}}) /R(T_0)$ numerically and determined $\alpha$ to best reproduce the experimental data. No other fit parameters were used. The results are shown in Fig.~\ref{fig:peak_heights_theory_fit} for different numbers of H$_2^+$ and H$_3^+$ ions. For our parameters the model predicts a nearly linear dependence of the scattering rate on the number of dark ions, which agrees with the experimental observations.

With a larger number of dark ions, the model predicts that the fluorescence peak height starts to saturate, and the fluorescence steps become smaller accordingly. We have conducted experiments for a larger number of dark ions and confirmed that up to eight dark ions can be reliably counted under our experimental conditions. Additional molecular dynamics simulations with different numbers of dark/coolant ions and different dark-ion species suggest that our method is robust for a wide range of experimental parameters.

In conclusion, we have demonstrated that the secular excitation technique can be used to detect dark ions embedded in laser-cooled ion Coulomb crystals with single-particle resolution. Secular excitation only relies on the Coulomb interaction between the different ion species. We therefore believe that this method could be used for single-particle sensitive detection of a wide range of atomic or molecular ions that do not possess suitable cycling transitions for fluorescence detection.

This project has received funding from the European Research Council (ERC) under the European Union’s Horizon 2020 research and innovation programme (grant agreement No. 742247). T.~W.~Hänsch acknowledges support from the Max Planck Foundation. The authors acknowledge the Max Planck Computing and Data Facility (MPCDF) for providing computing time.

\end{document}


\title{Supplemental Material for ``Number-Resolved Detection of Dark Ions in Coulomb Crystals''}
\date{\today}

\author{Fabian Schmid}
\email{fabian.schmid@mpq.mpg.de}
\affiliation{Max-Planck-Institut für Quantenoptik, 85748 Garching, Germany}
\author{Johannes Weitenberg}
\affiliation{Max-Planck-Institut für Quantenoptik, 85748 Garching, Germany}
\author{Jorge Moreno}
\affiliation{Max-Planck-Institut für Quantenoptik, 85748 Garching, Germany}
\author{Theodor W. Hänsch}
\affiliation{Max-Planck-Institut für Quantenoptik, 85748 Garching, Germany}
\affiliation{Fakultät für Physik, Ludwig-Maximilians-Universität München, 80799 München, Germany}
\author{Thomas Udem}
\affiliation{Max-Planck-Institut für Quantenoptik, 85748 Garching, Germany}
\affiliation{Fakultät für Physik, Ludwig-Maximilians-Universität München, 80799 München, Germany}
\author{Akira Ozawa}
\email{akira.ozawa@mpq.mpg.de}
\affiliation{Max-Planck-Institut für Quantenoptik, 85748 Garching, Germany}

\maketitle

\section{Molecular dynamics simulations}
We use the LAMMPS molecular dynamics code \cite{thompson2022, bentine2020} to simulate the secular excitation dynamics in our experiment. The ion trap is modelled as the sum of an oscillating and a static quadrupole electric field
\begin{equation}
    \mathbf{E}(\mathbf{r}, t) = \frac{U_\mathrm{rf} \cos(\Omega t) + U_\mathrm{dc}}{r_0^2} (y \mathbf{\hat{y}} - x \mathbf{\hat{x}}) + \kappa \frac{U_\mathrm{ec}}{2 z_0^2} (x \mathbf{\hat{x}} + y \mathbf{\hat{y}} - 2 z \mathbf{\hat{z}}),
    \label{eq:trapping_field}
\end{equation}
where $U_\mathrm{rf}$ and $\Omega$ are the amplitude and frequency of the trap radio frequency drive, $r_0$ is the radial size of the trap which depends on the distance between the trap electrodes and the trap center and on the electrode shape, $U_\mathrm{dc}$ is a static offset voltage between the two diagonal pairs of trap electrodes (see Fig.~1 in the main text), $U_\mathrm{ec}$ is the voltage applied to the endcap electrodes, $z_0$ is the distance between the endcap electrodes and the trap center, and $\kappa$ is a geometrical factor that depends on the trap geometry~\cite{berkeland1998}. $\mathbf{\hat{x}}$, $\mathbf{\hat{y}}$, and $\mathbf{\hat{z}}$ are orthogonal unit vectors that define the Cartesian coordinate system (see Fig.~1 in the main text). Table~\ref{tab:sim_parameters} shows the parameters used in our simulations.

\begin{table}[h]
    \centering
    \caption{Simulation parameters}
    \label{tab:sim_parameters}
    \begin{ruledtabular}
        \begin{tabular}{l l}
            Parameter & Value\\
            \hline
            $U_\mathrm{rf}$ & 121~V \\
            $\Omega$ & $2 \pi \times 66.05$~MHz \\
            $r_0$ & 0.469~mm \\
            $U_\mathrm{dc}$ & 0.1~V \\
            $U_\mathrm{ec}$ & 400~V \\
            $z_0$ & 3.5~mm \\
            $\kappa$ & 0.0469 \\
            $\Delta_1$ & $-2 \pi \times 130$~MHz \\
            $\Delta_2$ & $-2 \pi \times 460$~MHz \\
            $s_1$ & 1 \\
            $s_2$ & 1 \\
            $N_\mathrm{Be}$ & 2000 \\
        \end{tabular}
    \end{ruledtabular}
\end{table}

The lifetime of the excited state of the Be$^+$ cooling transition is 8.9~ns~\cite{Fuhr2010}. This is much shorter than the axial secular oscillation period of the Be$^+$ ions in the trap (1.6~µs). In this unresolved sideband regime, we model the laser cooling as a velocity-dependent average force~\cite{Leibfried2003} that only acts on the Be$^+$ ions
\begin{equation}
    \mathbf{F}_\mathrm{c}(v_z) = \sum_{j=1}^2 \hbar k \Gamma \frac{s_j/2}{1 + s_j + (\frac{\Delta_j - k v_z}{\Gamma/2})^2} \mathbf{\hat{z}},
    \label{eq:cooling_force}
\end{equation}
where $\Delta_j$ and $s_j$ are the detunings and the saturation parameters of the two cooling laser frequency components (see Table~\ref{tab:sim_parameters}), $\Gamma$ = $2\pi \times 18$~MHz is the FWHM linewidth of the Be$^+$ cooling transition~\cite{Fuhr2010}, $k = 2 \pi/(313~\mathrm{nm})$ is the cooling laser wave number, and $v_z$ is the $z$-component of the ion velocity.

The secular excitation is implemented as an oscillating homogeneous electric field
\begin{equation}
    \mathbf{E}_\mathrm{ex}(t) = E_\mathrm{ex} \cos(\Omega_\mathrm{ex} t) \frac{1}{\sqrt{2}}(\mathbf{\hat{x}} + \mathbf{\hat{y}}),
    \label{eq:secular_excitation_field}
\end{equation}
where $E_\mathrm{ex}$ and $\Omega_\mathrm{ex}$ are the amplitude and frequency of the excitation.

The fluorescence signal measured in the experiment is proportional to the rate of photons scattered by the Be$^+$ ions

\begin{equation}
    R(t) = \sum_{i=1}^{N_\mathrm{Be}} \sum_{j=1}^2 \Gamma \frac{s_j/2}{1 + s_j + (\frac{\Delta_j - k v_{z,i}(t)}{\Gamma/2})^2},
    \label{eq:scattering_rate}
\end{equation}

where $N_\mathrm{Be}$ is the number of Be$^+$ ions and $v_{z,i}$ is the $z$-component of the velocity of the $i$-th Be$^+$ ion.

\subsection{Secular excitation spectra}
For each number of H$_2^+$ and H$_3^+$ ions a simulated ion crystal is initialized as follows. First, we randomly place all ions in the simulation volume. Then, they are brought close to an equilibrium configuration by using the potential energy minimization routine of LAMMPS in a static pseudopotential as described in~\cite{bentine2020}. With that the H$_2^+$ and H$_3^+$ ions are located close to the trap axis because the pseudopotential is deeper for them than for the Be$^+$ ions. Finally, the pseudopotential is replaced by the trap electric field given in Eq.~(\ref{eq:trapping_field}) and laser cooling according to Eq.~(\ref{eq:cooling_force}) is applied to the Be$^+$ ions. The disordered secular motion of the many interacting ions in large Coulomb crystals behaves like the motion of a gas in thermal equilibrium and can therefore be assigned a temperature~\cite{zhang2007}. The laser cooling force damps this motion until the ions reach a secular temperature below 1~mK.

Then the secular excitation field given in Eq.~(\ref{eq:secular_excitation_field}) is turned on and the ion trajectories are simulated for 10~ms. The secular excitation heats up the secular motion until a thermal equilibrium is reached after a few ms. We apply an artificial damping force to the motion of the H$_2^+$ and H$_3^+$ ions during the first 3~ms of the simulation. This prevents the ion temperature from rising too quickly which occasionally leads to numerical instabilities in our simulation. The fluorescence signal observed in the experiment is simulated by averaging the Be$^+$ scattering rate calculated using Eq.~(\ref{eq:scattering_rate}) over the last 5~ms of the simulation. The simulation is then repeated with the same initial ion crystal for a range of secular excitation frequencies in order to obtain a secular spectrum. At typical secular temperatures reached in the experiment and simulations, the ions diffuse through the crystal on a ms time scale~\cite{zhang2007, roth2008a}. This provides enough averaging such that one run is sufficient for each combination of ion numbers and excitation frequencies. The resulting spectra for different numbers of H$_2^+$ and H$_3^+$ ions are shown in Fig.~5~(c) and (d) in the main text.

The all-to-all Coulomb interaction between the ions makes simulating large ion crystals computationally expensive. In our case, simulating the system evolution for 10~ms takes around one day on 8 CPU cores. We found that adding more cores did not speed up the simulation due to communication overhead between the cores. We therefore performed the simulations for the different excitation frequencies and ion numbers in parallel on a computer cluster. 

\subsection{Heating and cooling powers}
In Eq.~(3) in the main text we make two assumptions. The first one is that the total heating power due to the secular excitation is proportional to the number of H$_2^+$ or H$_3^+$ ions that are being resonantly excited. The second one is that the secular excitation is the only heating mechanism which is balanced by the laser cooling power in steady-state. The molecular dynamics simulation allows us to access such thermodynamic quantities of the system that cannot directly be measured in the experiment. The total heating power due to the excitation field is given by
\begin{equation}
    p_\mathrm{h,tot}(t) = \sum_{i=1}^{N_\mathrm{ion}} q_i \mathbf{v}_i(t) \cdot \mathbf{E}_\mathrm{ex}(t),
    \label{eqn:heating_power}
\end{equation}
where $N_\mathrm{ion}$ is the number of ions in the crystal, and $q_i$ and $\mathbf{v}_i(t)$ are the charge and velocity of the $i$-th ion.

The total laser cooling power is
\begin{equation}
    p_\mathrm{c,tot}(t) = -\sum_{i=1}^{N_\mathrm{Be}} \mathbf{v}_i(t) \cdot \mathbf{F}_\mathrm{c}(\mathbf{v}_i(t)).
    \label{eqn:cooling_power}
\end{equation}

\begin{figure}[ht]
    \centering
    \includegraphics[width=\linewidth]{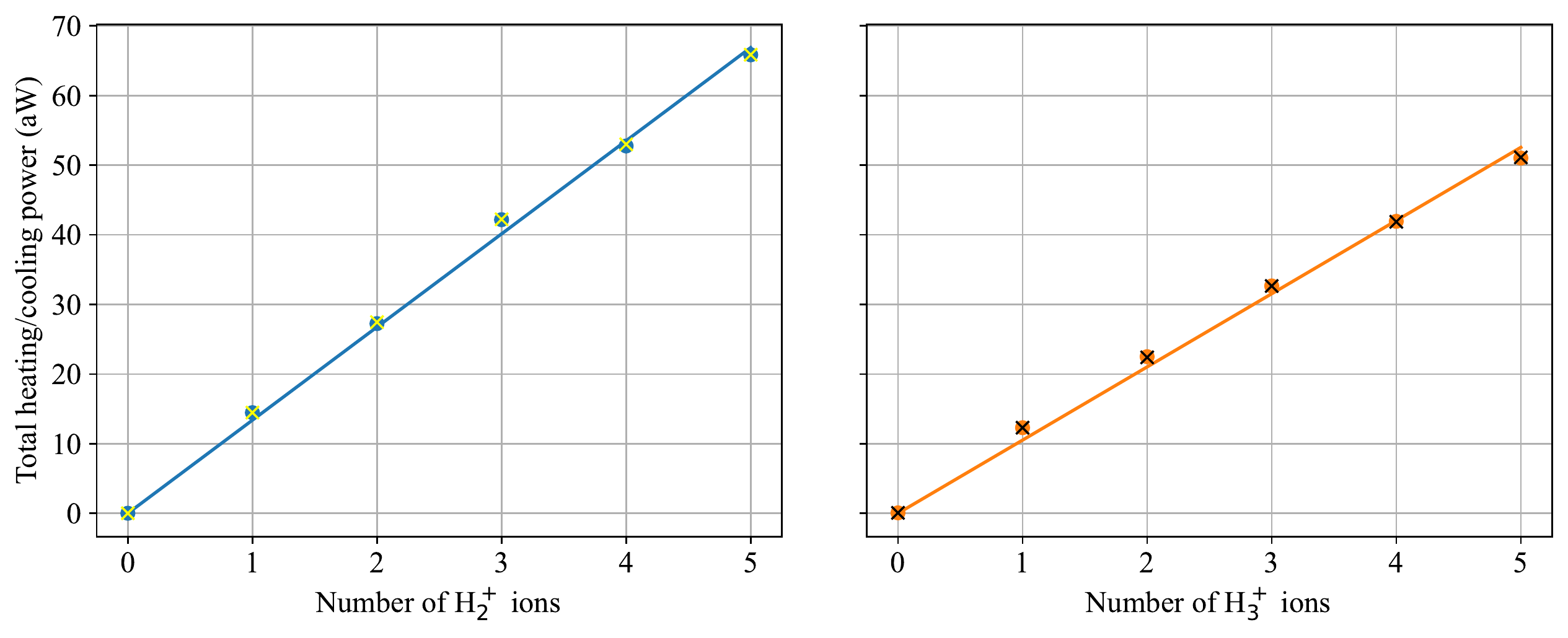}
    \caption{Secular heating powers (dots) and laser cooling powers (crosses) when exciting mixed ion crystals at the H$_2^+$ (left) and H$_3^+$ (right) resonance frequencies. The solid lines are linear fits to the heating powers.}%
    \label{fig:secular_heating_power}
\end{figure}

As for the scattering rate, we obtain steady-state values of the heating and cooling powers by averaging over the last 5~ms of each simulation. Fig.~\ref{fig:secular_heating_power} shows the heating and cooling powers during secular excitation at the H$_2^+$ and H$_3^+$ resonance frequencies for ion crystals with 2000 Be$^+$ ions and different numbers of H$_2^+$ and H$_3^+$ ions. The simulations confirm the approximate linear dependence between the heating power and number of ions. The simulations also show that the heating power is balanced by the laser cooling and that the trapping field does not transfer a significant amount of energy to the ions (RF heating~\cite{ryjkov2005}).